\documentclass[12pt]{article}
\usepackage[margin=3cm]{geometry}

\usepackage{amsmath}
\usepackage{amssymb}
\usepackage{graphicx}
\usepackage{subcaption}
\usepackage{bm,bbm}
\usepackage{xcolor}
\usepackage[normalem]{ulem}

\newcommand{\bra}[1]{\langle#1|}
\newcommand{\ket}[1]{|#1\rangle}
\newcommand{\braket}[2]{\langle#1|#2\rangle}

\newcommand{\comment}[1]{} 

\renewcommand{\t}[1]{\tilde{#1}}
\renewcommand{\dag}{^\dagger}
\newcommand{\Ord}{\mathcal{O}}

\newcommand{\C}{\mathbbm{C}}

\newcommand{\Ga}{\alpha}
\newcommand{\Gb}{\beta}
\newcommand{\Gc}{\gamma}

\newcommand{\Gs}{\sigma}
\newcommand{\Ge}{\epsilon}

\newcommand{\Gk}{\kappa}
\newcommand{\Gt}{\tau}

\newcommand{\GD}{\Delta}
\newcommand{\Gl}{\lambda}

\newcommand{\ve}[1]{\mathbf{#1}}

\newcommand{\nmin}{\textnormal{min}}

\newcommand{\nhhl}{\textnormal{HHL}}
\newcommand{\nvar}{\textnormal{var}}

\allowdisplaybreaks

\title{Error convergence of quantum linear system solvers}

\author{Matias Ginzburg$^a$, Ugo Marzolino$^{b,c}$ \\
\small{$^a$ SISSA, Italy} \\
\small{$^b$ University of Trieste, Italy} \\
\small{$^c$ INFN, Trieste Unit, Italy} \\
}
\date{}

\begin{document}

\newcommand{\e}[1]{ e^{\scalebox{0.7}{$\displaystyle #1$ }} }

\maketitle

\begin{abstract}
We analyze the performance of the Harrow-Hassidim-Lloyd algorithm (HHL algorithm) for solving linear problems and of a variant of this algorithm (HHL variant) commonly encountered in literature. This variant relieves the algorithm of preparing an entangled initial state of an auxiliary register. We prove that the computational error of the variant algorithm does not always converge to zero when the number of qubits is increased, unlike the original HHL algorithm. Both algorithms rely upon two fundamental quantum algorithms, the quantum phase estimation and the amplitude amplification.
In particular, the error of the HHL variant oscillates due to the presence of undesired phases in the amplitude to be amplified, while these oscillations are suppressed in the original HHL algorithm.
Then, we propose a modification of the HHL variant, by amplifying an amplitude of the state vector that does not exhibit the above destructive interference.
We also study the complexity of these algorithms in the light of recent results on simulation of unitaries used in the quantum phase estimation step,
and show that the modified  algorithm has smaller error and lower complexity than the HHL variant.
We supported our findings with numerical simulations.
\end{abstract}

\section{Introduction}

Quantum computers are machines made of physical systems ruled by the laws of quantum mechanics. The possibility to explore the whole Hilbert space allows for computational improvements over known classical algorithm \cite{nielsen2010quantum,BCRS2019}.
It is conjectured that quantum algorithms can solve in polynomial time (with respect to the problem size) problems than cannot be solved polynomially by classical algorithms, like the factorization problem \cite{Shor1994}, and even problems whose solutions are not thought to be polynomially verifiable by classical computers, like the boson sampling problem \cite{AaronsonArkhipov2011,AaronsonArkhipov2013} (namely outside the so-called NP class), but cannot probably solve NP-complete problems, i.e., the hardest problems in NP \cite{watrous2009,arnault2024typology}.

Among computational problems that are ubiquitous in computer science and big data \cite{SagirogluSinanc2013,FanHanLiu2014} and attract attention in quantum computation,
there are solutions of linear systems.
The inputs of a linear problem are a vector $\ve{b} \in \C^d$ with components $b_j$, $j=0,\dots d-1$ and a matrix $A \in \C^{d \times d}$ with components $A_{i,j}$, eigenvectors $\ve{u}_j$ and eigenvalues $\Gl_j$, $i,j=0,\dots d-1$. The output is a vector $\ve{x} \in \C^d$ satisfying
\begin{equation}\label{eq:Ax=b}
    A\cdot\ve{x}=\ve{b} \,.
\end{equation}
In the following analysis, we consider $s$-sparse matrices, namely matrices with a maximum number $s$ of non-vanishing elements in each row or column, and define the condition number as $\Gk=\max_j|\Gl_j|/\min_j|\Gl_j|$.

The understanding of the potentialities of quantum algorithms for linear problems is then of great importance for several potential applications.
The first quantum algorithm for linear systems was derived by Harrow, Hassidim and Lloyd \cite{HHL_linear_equations} (henceforth HHL algorithm).
The HHL algorithm exploits two fundamental subroutines, namely the quantum phase estimation \cite{kitaev1995quantum,Cleve_1998,AbramsLloyd1999,nielsen2010quantum} and the amplitude amplification \cite{Brassard_2002} (a variant of the Grover's search algorithm \cite{grover1996fast,Grover_1997,Boyer_1998}) that represent the core for general quantum algorithms (solving the so-called BQP-complete problems) \cite{Cade2023}.


In literature, the HHL algorithm is sometimes applied using a simpler, fully factorized initial state of an auxiliary quantum register \cite{cao2012quantum,barz2013solving, Pan_2014,dervovic2018quantum,Duan2020,morrell2023stepbystep}.
This simplification allows to reduce the complexity of the initial state preparation that is entangled in the original HHL algorithm.
In the following, we refer to this version of the HHL algorithm as \emph{HHL variant}.
These algorithms are also used as subroutines of more complicated algorithms, like those for solving linear and non-linear differential equations \cite{jin2022quantumObservables,ito2023map}.
The simpler state preparation of the HHL variant is advantageous
to design near-term experimental realisations
with noisy intermediate-scale quantum (NISQ) technologies \cite{Preskill2018,Bharti2022}.

Within this framework, it is important to understand how the performances of these algorithms scale with the amount of resources, namely the number of qubits and the number of gates.
We show that the computational error of the HHL variant does not approach zero for large registers when the condition numbers $\Gk$ is large or using a poor estimation of $\Gk$.
With minimal modifications to the algorithm structure,
we derive an improved algorithm which provides solutions up to small errors with a lower complexity than the HHL variant.
We also show that the original HHL algorithm does not suffer from the above convergence issue and always provides better performances than the HHL variant. The different convergence behaviour relies on the aforementioned state preparation that is simpler in the HHL variant.

Our analysis then sheds light on the theoretical structure of quantum algorithms for linear problems, and can be generalized to algorithms exploiting recently developed alternatives the quantum phase estimation \cite{Higgins2009,SvoreHastingsFreedman2013,Rall2021,LinTong2022,DongLinTong2022,Kurdzialek2023,Patel2024}.
We also carefully consider the complexity of the algorithm in the light of recent results on simulations of unitaries \cite{berry2007efficient,BerryChilds2012,
BerryCleveGharibian2014,Berry2014,Berry_2015,Berry2015-2,BerryNovo2016,BerryNovo2017,
Low_2017,Low_2019} used in the quantum phase estimation step.

In section \ref{sec:HHL}, we review the original HHL algorithm and its variant. In section \ref{sec:ErrorAnalysis}, we analyse the convergence of the algorithm outcome towards the exact solution. The improved HHL variant is derived in section \ref{sec:new}. We compare the complexity of the above algorithms in section \ref{sec:compl}. Sections \ref{sec:num} and \ref{sec:neg} are devoted, respectively, to numerical simulations of the HHL variant and the improved algorithm, and to the relaxation of a technical assumption (the positivity of $A$). We draw conclusions in section \ref{sec:concl}, and report some technical and lengthy derivations in appendices.

\section{HHL algorithm} \label{sec:HHL}

The linear problem \eqref{eq:Ax=b} has a solution if and only if $\ve{b}$ lies in the support of $A$. In this case, given the eigenvectors $\ve{u}_j$ of $A$ with eigenvalues $\Gl_j\neq 0$, there is a decomposition of the input vector $\ve{b} = \sum_j \Gb_j \ve{u}_j$ for some $\Gb_j\in \C$ and the solution is $ \ve{x} = \sum_j (\Gb_j /\Gl_j) \ve{u}_j$. An alternative decomposition of the input state is $\ve{b} = \sum_j b_j \ve{e}_j$, where $\ve{e}_j$ are basis vectors with elements $\big(\ve{e}_j\big)_l=\delta_{j,l}$.

For later convenience, assume that $\ve{b}$ has unit $L^2$-norm, $\|\ve{b}\|_2=1$. If this condition is not met, the linear problem \eqref{eq:Ax=b} can be mapped to an equivalent problem by dividing $\ve{b}$ and $\ve{x}$ by the norm $\|\ve{b}\|_2$.
Assume also that $A$ is Hermitian, as the problem with non-Hermitian $A$ can be embedded in $\C^{2d}$ with an Hermitian matrix.
Consider indeed the following problem
\begin{equation}
\begin{pmatrix}
0 & A \\
A^\dagger & 0
\end{pmatrix}
\cdot\ve{y}=
\begin{pmatrix}
\ve{b} \\
\ve{0}
\end{pmatrix}\,,
\end{equation}
whose solution is the solution of the original problem embedded in $\C^{2d}$: $\ve{y}=\ve{0}\oplus A^{-1}\ve{b}=\ve{0}\oplus\ve{x}$ where $\ve{0}\in\C^d$ is the vector with all components set to zero. Solving the new problem, then solves the original problem at the cost of doubling the dimension of the solution space.
Finally, assume without loss of generality that the eigenvalues of $A$ satisfy $0<\Gl_j\leqslant 1$. The conditions $\Gl_j\leqslant 1$ can be achieved by dividing $A$ and multiplying $\ve{x}$ by a rough estimate (rounded up) of the largest eigenvalue of $A$.
We will show in Section \ref{sec:neg} how to relax the condition $\Gl_j>0$.

The HHL algorithm and its variant use three quantum registers.
\begin{itemize}
\item
The register $r$ consists of $n=\lceil \log_2 d \rceil$ qubits used to encode the vector $\ve{b}$ in the pure state $\ket{b}_r = \sum_{j=0}^{d-1} b_j \ket{j}_r$, where $\{|j \rangle_r\}_{j=0,\dots,d-1}$ is the computational basis, i.e. the eigenvectors of the Pauli matrix $\sigma_z$ for each qubit.
This is the reason for requiring $1=\|\ve{b}\|_2^2=\langle b|b\rangle$.
This pure state can also be expressed as $\ket{b}_r=\sum_j \Gb_j \ket{u_j}_r$, where $\ket{u_j}_r=\sum_k V_{k,j}\ket{j}_r$ and $V=[V_{k,j}]$ is the unitary matrix that diagonalises $A$, $V^{\dag}A\,V=\textnormal{diag}\,(\lambda_1,\dots,\lambda_d)$.
\item
The register $c$, called \emph{clock}, is used to implement the quantum phase estimation and
to store the binary representation of eigenvalues of $A$
as elements of the
computational basis.
It consists of $n_c$ qubits, so that the states in the computational basis represent up to $T=2^{n_c}$ different values and consequently
the maximum precision for eigenvalue estimations is $2^{-n_c}=T^{-1}$.
When $n_c$ and $T$ increase, the eigenvalues of $A$ are discriminated with better precision leading to a smaller error in the final output.
For this reason, we use $T$ as the parameter to analyse the convergence towards the exact solution.
\item
The algorithm also uses an additional ancillary qubit $a$.
\end{itemize}

The solution $\ve{x}$ is encoded in the amplitudes of the output state vector of the $r$-register. Nevertheless, it is impossible to generate a state whose amplitudes are the components of $\ve{x}$, namely $\sum_jx_j\ket{u_j}_r$ with $x_j=\beta_j/\lambda_j$, because it is not normalized in general.
Instead, the HHL algorithm aims at producing a state as close as possible to $\sum_jx_j\ket{u_j}_r/\|\ve{x}\|_2$.
Unfortunately, the amplitudes can be precisely measured with state tomography which is not efficient in the worst case \cite{ParisRehacek2004,GrossLiuFlammiaBeckerEisert2010,Gross2011,flammia2012quantum}. Therefore, the HHL algorithm does not compute the exact solution $\ve{x}$, but it can be used to compute the norm $\|\ve{x}\|_2$ and expectation values $\ve{x}\dag M \ve{x}$
of matrices $M$
that can be implemented efficiently on the quantum register.

\begin{figure}[t!]
    \centering
    \includegraphics[width = 0.8\textwidth]{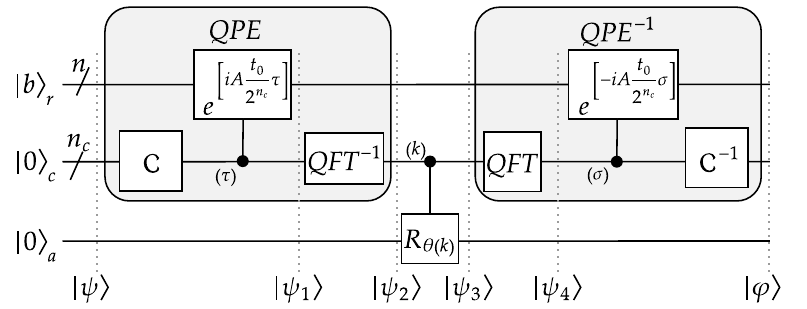}
    \caption{Circuit of the main part of the HHL algorithm and of its variant.
    The first three gates implement the phase estimation protocol. The operation $\mathcal{C}$ represents the preparation of the \textit{clock}, it can be chosen as in \eqref{eq:clock.HHL} to implement the original algorithm or as in \eqref{eq:clock.var} to implement its variant.  The notations $(\Gt)$, $(k)$ and $(\Gs)$ indicate controlled operations conditioned on the computational basis states $\ket{\Gt}_c$, $\ket{k}_c$ and $\ket{\Gs}_c$ of the $c$-register respectively. The angle of the rotation $R_{\theta(k)}$ are defined by
    $\sin{\theta(k)} := Ct_0/(2\pi k)$. The last three steps uncompute the phase estimation.}
    \label{fig:HHL_schematic}
\end{figure}

The main part of the original HHL algorithm \cite{HHL_linear_equations} and of its variant \cite{cao2012quantum,barz2013solving, Pan_2014,dervovic2018quantum,Duan2020,morrell2023stepbystep} is compactly described by the circuit in figure \ref{fig:HHL_schematic}.
The explicit formulas of intermediate states are reported in Appendix \ref{app:steps}.
The quantum registers
are initially prepared in the following state
\begin{equation}
\ket{\psi} = \ket{0}_a\ket{0}_c\sum_j \Gb_j \ket{u_j}_r \,. \label{eq:psi_0}
\end{equation}
The difference between the HHL algorithm and its variant is the first operation on the clock, that is respectively
\begin{align}
    \mathcal{C}^\nhhl\ket{0}_c
	& = \sqrt{\frac{2}{T}}\,\sum_{\tau=0}^{T-1}\sin\left(\frac{\pi\left(2\,\tau+1\right)}{2\,T}\right)\ket{\tau}_c \,, \label{eq:clock.HHL} \\
    \mathcal{C}^\nvar\ket{0}_c
    & = \frac{1}{\sqrt{T}}\,\sum_{\tau=0}^{T-1} \ket{\tau}_c\,, \label{eq:clock.var}
\end{align}
In the HHL variant, the operator $\mathcal{C}^\nvar$ is the Hadamard gate on each qubit of the clock
$\ket{\Gc}_c\to H^{\otimes n_c}\ket{\Gc}_c=\frac{1}{\sqrt{T}}\sum_{\Gs=0}^{T-1} \exp\big(i\pi \sum_{l=1}^{n_c}\Gc_l\,\Gs_l\big) \ket{\Gs}_c$ with $\{\Gc_l\}_l$ and $\{\Gs_l\}_l$ the binary representations of $\Gc$ and $\Gs$.
Therefore, these Hadamard gates can be parallelized such that they require constant time, and the state $\mathcal{C}^\nvar\ket{0}_c$ is fully factorized.
The operator $\mathcal{C}^\nhhl$ in the HHL algorithm is more complicated as it entangles the clock qubits, and can be implemented in time $\Ord(n_c)=\Ord(\log_2T)$ \cite{kaye2001quantum,grover2002creating}.
We will show later that this small complexity reduction of the HHL variant goes with a substantial reduction of the precision of eigenvalue estimations.
The initial state of the $r$-register can be prepared following known state preparation schemes \cite{Grover2000,kaye2001quantum,grover2002creating,Mottonen_states,Soklakov2006,
Plesch2011,Araujo2021,Zhang2021,wang2021fast,bausch2022fast,clader2022quantum,
low2024trading}, or may result from a previous quantum protocol.

In figure \ref{fig:HHL_schematic}, we have denoted by
$QFT$ the quantum Fourier transform
$\ket{\Gt}_c\to \frac{1}{\sqrt{T}}\sum_{k=0}^{T-1} \e{-i2\pi{\Gt k}/{T}} \ket{k}_c$,
and by $QFT^{-1}$ its inverse. The other gates are controlled unitaries defined by the following state transformations
\begin{align}
\label{eq:CexpA}
\ket{\tau}_c\ket{b}_r
& \longrightarrow\ket{\tau}_c\,\e{iA\frac{t_0}{T}\tau}\ket{b}_r \,, \\
\label{eq:conditionated_Rotation}
\ket{0}_a\ket{k}_c
& \longrightarrow
\begin{cases}
\sin\big(\theta(k)\big)\,\ket{1}_a\ket{k}_c + \cos\big(\theta(k)\big)\,\ket{0}_a\ket{k}_c
& \textnormal{if } k\geqslant k_\nmin \\
\ket{0}_a\ket{k}_c
& \textnormal{if } k<k_\nmin \\
\end{cases} \,,
\end{align}
where $t_0$ and $k_\nmin$ are free parameters
and $\sin\big(\theta(k)\big) := C t_0/(2\pi k)$ with $0<C\leqslant {2\pi k_\nmin}/{t_0}$.
The implementability of the transformation \eqref{eq:CexpA} on a quantum computer requires the Hermiticity of $A$ assumed above.

The state vector after these operations for both algorithms is of the form
\begin{align}
\ket{\varphi}
& =\frac{1}{\sqrt{T}} \sum_{j=0}^{2^{n}-1} \sum_{\Gs=0}^{T-1} \sum_{k=k_\nmin}^{T-1} \Gb_j\, \Ga_{k|j}\, \sin\big(\theta(k)\big)
\,\e{i \frac{\Gs}{T}(2\pi k-\Gl_j t_0)}
\ket{1}_a\,\mathcal{C}^{-1}\ket{\Gs}_c\ket{u_j}_r \nonumber \\
& +\frac{1}{\sqrt{T}} \sum_{j=0}^{2^{n}-1} \sum_{\Gs=0}^{T-1} \sum_{k=k_\nmin}^{T-1} \Gb_j\, \Ga_{k|j}\, \cos\big(\theta(k)\big)
\,\e{i \frac{\Gs}{T}(2\pi k-\Gl_j t_0)} \ket{0}_a\,\mathcal{C}^{-1}\ket{\Gs}_c\ket{u_j}_r \nonumber \\
& +\frac{1}{\sqrt{T}} \sum_{j=0}^{2^{n}-1} \sum_{\Gs=0}^{T-1} \sum_{k=0}^{k_\nmin-1} \Gb_j\, \Ga_{0|j}
\,\e{i\frac{\Gs}{T}(2\pi k-\Gl_j t_0)}
\ket{0}_a\,\mathcal{C}^{-1}\ket{\Gs}_c\ket{u_j}_r
\,,
\label{eq:psi_7}
\end{align}
with coefficients
\begin{align}
\Ga_{k|j}^\nhhl
& =  \frac{\sqrt{2}}{T} \sum_{\tau=0}^{T-1} \e{i\frac{\Gt}{T}(\Gl_j t_0-2\pi k)}
\sin\left(\frac{\pi\left(2\,\tau+1\right)}{2\,T}\right) \nonumber \\
& = \e{i\phi} \frac{\sqrt{2}\cos\left(\frac{1}{2}(\Gl_j t_0 -2\pi k)\right) \cos\left(\frac{1}{2T}(\Gl_j t_0 -2\pi k)\right) \sin\left(\frac{\pi}{2T}\right)}{T \sin\left(\frac{1}{2T}(\Gl_j t_0 -2\pi k + \pi)\right) \sin\left(\frac{1}{2T}(\Gl_j t_0 -2\pi k - \pi)\right)} \,,
\label{eq:alpha.HHL} \\
\Ga_{k|j}^\nvar
& =  \frac{1}{T} \sum_{\tau=0}^{T-1} \e{i\frac{\Gt}{T}(\Gl_j t_0-2\pi k)} = \e{i\phi} \frac{\sin(\frac{1}{2}(\Gl_j t_0 - 2\pi k))}{T\sin(\frac{1}{2T}(\Gl_j t_0 - 2\pi k)) } \,,
\label{eq:alpha}
\end{align}
for the original HHL algorithm and its variant.
The phase $\phi=\frac{1}{2}(\Gl_j t_0 -2\pi k)\big(1-\frac{1}{T}\big)$ is not relevant in the following analysis.

The first three gates in figure \ref{fig:HHL_schematic} for the HHL variant are the standard quantum phase estimation algorithm \cite{kitaev1995quantum,Cleve_1998,AbramsLloyd1999,nielsen2010quantum} used to estimate the eigenvalues of $A$.
The different operation $\mathcal{C}^\nhhl$ in the original HHL algorithm modifies this subroutine which however serves the same purposes with improved estimation precision \cite{LuisPerina1996,BuzelDerkaMassar1999,BerryWiseman2000} and can be also termed as quantum phase estimation.
The controlled rotation \eqref{eq:conditionated_Rotation} entangles the $r$- and the $c$-register with the ancillary qubit that plays the role of a flag for an approximated solution in the expansion \eqref{eq:psi_7}.
Indeed,
since the coefficients $\Ga_{k|j}$ are centered around ${2\pi k}/{t_0} \approx \lambda_j$
and recalling $\sin{\theta(k)} := Ct_0/(2\pi k)$,
the coefficients in the first line of equation \eqref{eq:psi_7} are estimates of the desired coefficients $x_j=\beta_j/\lambda_j$.

A wise choice of $k_\nmin$ guarantees that the $r$-register output is entangled with the clock states that reliably store all the eigenvalues.
Indeed, the controlled rotations \eqref{eq:conditionated_Rotation} should be applied to the indices $k\geqslant k_\nmin$ such that the coefficients $\Ga_{k|j}$ in the first line of equation \eqref{eq:psi_7} cover all their peaks around ${2\pi k}/{t_0} \approx \lambda_j$ for all $\Gl_j$.
This happens when $k_\nmin<\,t_0\,\Gl_\nmin/(4\pi)$
and $\Gl_\nmin=\min_j\Gl_j$ is the minimum eigenvalue of $A$. If an upper bound of the condition number $\kappa\leqslant\kappa'$ is available, then we set $k_\nmin=\lfloor t_0/(4\pi\kappa')\rfloor$.
Therefore, $t_0$, $k_\nmin$ and $\Gk'$ are parameters of the algorithm that can be chosen in order to minimize the computational error and the complexity of the algorithm, as we shall discuss later.
In practice, the computation of the condition number may be time consuming, or estimates $\kappa'$, as well as $\kappa$ itself, can be large for hard problems.
Without any information on the condition number, we need to set $k_\nmin=1$
which corresponds to an effective bound $\kappa'=t_0/(4\pi k_\nmin)=t_0/(4\pi)$.
The value $\kappa'=t_0/(4\pi)$ is then an upper bound of the condition number of $A$ in linear problems solvable by the HHL algorithm and its variant using $n_c$ qubits in the clock.
Applications of the HHL variant usually assume $k_\nmin=1$ \cite{cao2012quantum,barz2013solving, Pan_2014,dervovic2018quantum,Duan2020,morrell2023stepbystep}.
We shall discuss later the impact of large condition numbers on the computational error and on the complexity of the algorithms.

The component in the first line of the state \eqref{eq:psi_7} is the one that stores the approximated solution, and is the only one where the state $|1\rangle_a$ appears.
The outputs of both the original HHL algorithm and its variant are then achieved by measuring the ancillary qubit in the basis $\{|0\rangle_a,|1\rangle_a\}$ and postselecting the outcome $1$ which occurs with probability
\begin{equation} \label{eq:ptilde}
p_0 = \sum_{j=0}^{2^{n}-1} |\Gb_j|^2 \, \sum_{k=k_\nmin}^{T-1} \big|\Ga_{k|j}\big|^2 \left( \frac{C\,t_0}{2\pi k}\right)^2 \,,
\end{equation}
which can be small.
Nevertheless, one can find the outcome $\ket{1}_a$ after applying the amplitude amplification algorithm \cite{Boyer_1998,Brassard_2002} to the state \eqref{eq:psi_7} with an average running time $\Ord\big(1/\sqrt{p_0}\big)$.
The final normalized outputs of the circuits in Figure \ref{fig:HHL_schematic} are of the form
\begin{equation} \label{eq:out}
\ket{x_0} = \frac{1}{\sqrt{T\,p_0}}\sum_{j=0}^{2^{n}-1} \sum_{\Gs=0}^{T-1} \sum_{k=k_\nmin}^{T-1} \Gb_j \, \Ga_{k|j} \, \frac{C\,t_0}{2\pi k}
\,\e{i \frac{\Gs}{T}(2\pi k-\Gl_j t_0)}
\ket{1}_a\,\mathcal{C}^{-1}\ket{\Gs}_c\ket{u_j}_r \,,
\end{equation}
with the coefficients \eqref{eq:alpha.HHL} or \eqref{eq:alpha}.

\section{Error analysis} \label{sec:ErrorAnalysis}

In this section, we discuss how close the output states of the HHL algorithm and of its variant are to the ideal solution of the linear problem \eqref{eq:Ax=b}.
The ideal output of the algorithm is the embedding of the vector $\ve{x}$ in the $r$-register as follows
\begin{equation} \label{x.id}
\ket{x}
= \ket{1}_a \ket{0}_c\,\frac{C}{\sqrt{p}}\,\sum_{j=0}^{2^{n}-1} \frac{\Gb_j}{\Gl_j}\,\ket{u_j}_r \,,
\end{equation}
where we wrote the normalization constant as $C/\sqrt{p}$ in order to make the expression more similar to \eqref{eq:out}, and the normalization of the state \eqref{x.id} entails
\begin{equation}
p = \sum_{j=0}^{2^{n}-1} \left( \frac{C\,|\Gb_j|}{\Gl_j}\right)^2 \,.
\end{equation}
Since $\sin\big(\theta(k)\big)=C t_0/(2\pi k)$ is an amplitude of the state in \eqref{eq:conditionated_Rotation}, the constant $C$ satisfies $C\leqslant2\pi k/t_0$ for all $k\geqslant k_\nmin$.
We then choose the largest of the allowed values $C=2\pi k_\nmin/t_0\simeq1/(2\kappa')$ that maximizes the ideal and the actual probability of measuring $\ket{1}_a$, respectively $p$ and $p_0$.

Before proceeding with our analysis, we stress that
the precision of the eigenvalue estimation
in the phase estimation subroutine is improved by increasing $t_0$. The reason is that the eigenvalues $\lambda_j$ are amplified and can be better distinguished. On the other hand, the phase estimation cannot differentiate phases modulo $2\pi$, and if $\Gl_j{t_0}/{T}>2\pi$ the protocol fails.
Therefore, under the assumption $0<\Gl_j\leqslant 1$ for all $j$, we can take $t_0$ proportional to $T$, $t_0= t T$ for some $0<t<2\pi$.
This scaling implies $p=\Ord(C/\Gl_\nmin)^2=\Ord\big((2\pi k_\nmin)^2/(\Gl_\nmin t T)^2\big)=\Ord(\kappa/\kappa')^2$.

The states $\ket{x}$ and $\ket{x_0}$ are close to each other if their squared overlap, also called fidelity $|\braket{x}{x_0}|^2$, is close to $1$. We then consider the so-called \emph{infidelity} $1-|\braket{x}{x_0}|^2$ as a measure of the deviation of the output state $\ket{x_0}$ from the ideal state $\ket{x}$.
Indeed, the square root of the infidelity satisfies the properties of a distance between pure states $d\big(\ket{x},\ket{x_0}\big)=\sqrt{1-|\braket{x}{x_0}|^2}$ \cite{Rastegin2002,Rastegin2006},
and we shall use $d\big(\ket{x},\ket{x_0}\big)$ to quantify the error of the algorithm.
A straightforward computation gives
\begin{align}\label{eq:Fid_x_out}
\braket{x}{x_0} & = \frac{C^2}{\sqrt{p \, p_0}} \, \sum_{j=0}^{2^{n}-1} \frac{|\Gb_j|^2}{\Gl_j} \, \sum_{k=k_\nmin}^{T-1} \big|\Ga_{k|j}\big|^2\,\frac{t\,T}{2\pi k} \nonumber \\
& = \frac{\displaystyle \sum_{j=0}^{2^{n}-1} \frac{|\Gb_j|^2}{\Gl_j} \, \sum_{k=k_\nmin}^{T-1} \big|\Ga_{k|j}\big|^2\,\frac{t\,T}{2\pi k}}
{
\sqrt{
\displaystyle \sum_{l=0}^{2^{n}-1} \frac{|\Gb_l|^2}{\Gl_l^2} \,
\sum_{j=0}^{2^n-1} |\Gb_j|^2 \, \sum_{k=k_\nmin}^{T-1} \left( \big|\Ga_{k|j}\big|\,\frac{t\,T\,}{2\pi k}\right)^2}
}\,.
\end{align}

Moreover, the following quantities
\begin{align}
\Ge_{1,j} & = \Gl_j\,\left(\sum_{k=k_\nmin}^{T-1} \big|\Ga_{k|j}\big|^2 \,\frac{t T}{2\pi k} - \frac{1}{\Gl_j}\right) \,, \label{epsilon1} \\
\Ge_{2,j} & = \Gl_j^2\,\left(\sum_{k=k_\nmin}^{T-1} \left|\Ga_{k|j}\right|^2  \left(\frac{t T}{2\pi k}\right)^2 - \frac{1}{\Gl_j^2}\right) \,, \label{epsilon2}
\end{align}
are useful to study the convergence of the distance $d\big(\ket{x},\ket{x_0}\big)$ for large dimension of the clock Hilbert space, $T=2^{n_c}$,
\begin{equation} \label{eq:err.HHL}
d\big(\ket{x},\ket{x_0}\big)=
\sqrt{
\frac{\displaystyle
\sum_{j,l=0}^{2^n-1} \frac{|\Gb_j|^2\,|\Gb_l|^2}{\Gl_j^2\,\Gl_l^2}\left(\Ge_{2,j}-\Ge_{1,j}-\Ge_{1,l}-\Ge_{1,j}\,\Ge_{1,l}\right)
}
{\displaystyle
\sum_{j,l=0}^{2^n-1} \frac{|\Gb_j|^2\,|\Gb_l|^2}{\Gl_j^2\,\Gl_l^2} \left(1+\Ge_{2,j}\right)
}
} \,.
\end{equation}
The algorithm provides reliable solutions of linear systems only if $d\big(\ket{x},\ket{x_0}\big)$ approaches zero, namely for small $\Ge_{1,j}$ and $\Ge_{2,j}$.

\begin{figure}[t!]
    \centering
    \includegraphics[width = 0.45 \textwidth]{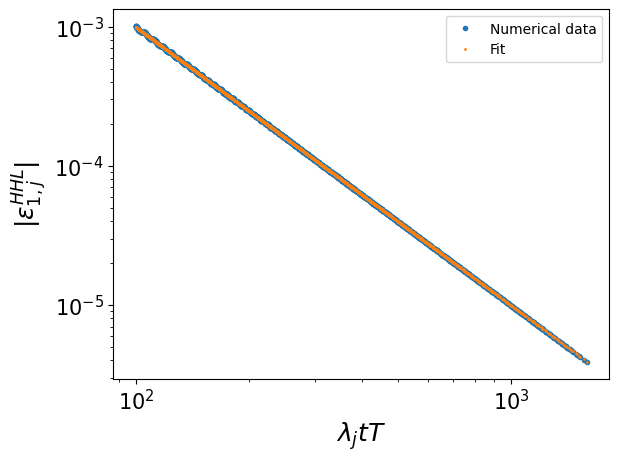}
    \includegraphics[width = 0.45 \textwidth]{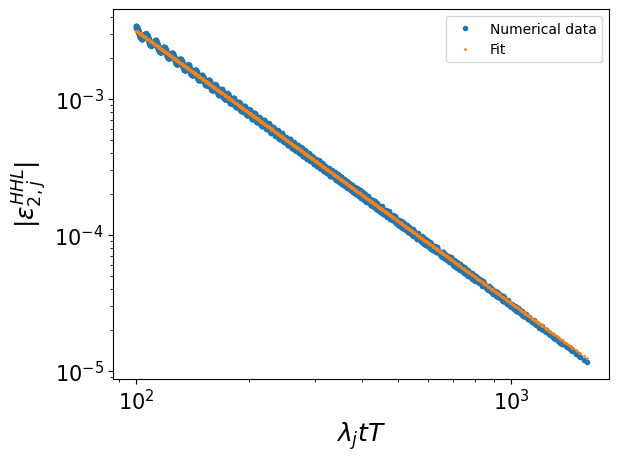}
    \caption{Numerical computations of $\Ge_{1,j}$ (left panel) and $\Ge_{2,j}$ (right panel) for $50$ equally spaced values of $\Gl_j\in(0,1)$, $50$ equally spaced values of $t\in(0.1\,\pi,\pi)$, and $n_c$ ranging from $3$ to $9$ ($T\in[2^3,2^9]$). The orange curves are fits with functions $a_1(\Gl_jtT)^{-2}$ (left panel) and $a_2(\Gl_jtT)^{-2}$ (right panel) which result in $a_1\simeq9.94$ and $a_2\simeq31.54$.}
    \label{fig:fit}
\end{figure}

We have computed numerically $\Ge_{l,j}^\nhhl$ ($l\in\{1,2\}$) for the original HHL algorithm for $50$ equally spaced values of $\Gl_j\in(0,1)$, $50$ equally spaced values of $t\in(0.1\,\pi,\pi)$, and
$n_c$ ranging from $3$ to $9$ ($T\in[2^3,2^9]$).
We also assume $k_\nmin=1$ that is the worst choice as higher values allow for shorter times of the amplitude amplification $\Ord\big(1/\sqrt{p_0}\big)$ which
decreases with increasing $C=2\pi k_\nmin/t_0$.
The numerical data are then fitted with the function $a_l(\Gl_jtT)^{-2}$. The fits, shown in figure \ref{fig:fit},
result in $a_1\simeq9.94$ and $a_2\simeq31.54$ with a very good agreement.
Given $\Gl_j\geqslant\Gl_\nmin\geqslant 1/\Gk$, the error of the orignal HHL algorithm is
\begin{equation} \label{eq:errHHL}
d\big(\ket{x},\ket{x_0^\nhhl}\big)\propto\frac{\Gk}{T} \,.
\end{equation}

For the HHL variant, $\Ge_{1,j}^\nvar$ decreases with $T$ slower than $\Ge_{1,j}^\nhhl$. In appendix \ref{app:bound1} we compute the following bound
\begin{equation} \label{eq:Gl-convergence}
\big|\Ge_{1,j}^\nvar\big| 
\leqslant \Ord\left(\frac{\ln(\Gl_j \,t \,T)}{\Gl_j \,t \,T}\right) \,,
\end{equation}
whose scaling with $\Gl_jtT$ is saturated by the exact numerical computations.
On the other hand,
we prove in appendix \ref{app:bound2} that the convergence of $\Ge_{2,j}^\nvar$ for the HHL variant depends on $k_\nmin$ (and therefore $\kappa'$),
\begin{equation}\label{eq:Gl_square}
\sin^2\left( \frac{\Gl_jtT}{2}\right) \left(\frac{2\kappa'}{tT} \right)^2
+ \Ord\left( \frac{\ln(\Gl_j t T)}{\Gl_j t T} \right)
\leqslant\big|\Ge_{2,j}^\nvar\big| \leqslant
\Ord\left(\frac{\kappa'}{tT}\right)
+
\Ord\left( \frac{\ln(\Gl_j t T)}{\Gl_j t T} \right) \,.
\end{equation}
Recalling $\Gl_j\geqslant\Gl_\nmin\geqslant 1/\Gk$, the lower bound of the computational error for the HHL variant is then
\begin{equation}\label{eq:error0}
d\big(\ket{x},\ket{x_0^\nvar}\big)
\geqslant
\begin{cases}
\displaystyle \Ord \left( \sqrt{\frac{\kappa}{T}\,\ln\frac{T}{\kappa}} \right)
& \displaystyle \textnormal{if } \Gk'=\Ord\left(\sqrt{\Gk\,T\,\ln\left(\frac{T}{\Gk}\right)}\right) \\
\displaystyle \sin^2\left( \frac{\Gl_jtT}{2}\right) \left(\frac{2\kappa'}{tT} \right)^2
& \displaystyle \textnormal{if } \Gk'\gg\Ord\left(\sqrt{\Gk\,T\,\ln\left(\frac{T}{\Gk}\right)}\right)
\end{cases}
\,.
\end{equation}
The inequalities \eqref{eq:Gl_square} and \eqref{eq:error0} imply that the HHL variant is accurate when both the condition number $\kappa$ and its computable bound $\kappa'$ are much smaller than $T$, resulting in $k_\nmin=\lfloor tT/(4\pi\kappa')\rfloor\gg1$.
The HHL variant is then not accurate
if the bound $\kappa'$ is as large as $T$ ($k_\nmin\sim1$) even when $\kappa\ll T$.


The convergence problem of the HHL variant can be understood from equation \eqref{eq:out}. Recalling that $\mathcal{C}^\nvar=H^{\otimes n_c}$ in this case, the coefficients of the states with fixed $\ket{\Gc}_c$ are proportional to
\begin{equation}
\frac{\alpha_{k|j}^\nvar}{T}\,\sum_{\Gs=0}^{T-1} \e{i\frac{\Gs}{T}(2\pi k -\Gl_j t_0 ) +i\pi\sum_{l=1}^{n_c} \Gs_l \Gc_l}
\end{equation}
which, for $\Gc=0$ and from equation \eqref{eq:alpha}, is equal to $\big|\Ga_{k|j}\big|^2$, and is centered at $k \approx \Gl_j t_0/(2\pi)$. Nevertheless, when $\Gc\neq0$, the phase is shifted and the final distribution is not centered around the correct value.
This problem is avoided by measuring the clock in the computational basis and postselecting the outcome $\Gc=0$, as discussed in the next section.

\section{Improved HHL variant} \label{sec:new}

We have shown that the HHL variant is not useful when the condition number is poorly estimated.
An improved HHL variant, that is accurate event without any bound $\kappa'$,
can be implemented with the amplitude amplification \cite{Boyer_1998,Brassard_2002} of the component $\ket{1}_a\ket{0}_c$, instead of $\ket{1}_a$, after the circuit that leads to the state \eqref{eq:psi_7}.
The probability $\t p$ of measuring $\ket{1}_a\ket{0}_c$ from the state \eqref{eq:psi_7} is
\begin{equation}
    \t p = \sum_j |\Gb_j|^2 \left(\sum_{k=k_\nmin}^{T-1} \big|\Ga_{k|j}^\nvar\big|^2 \, \frac{Ct T}{2\pi k}\right)^2.
\end{equation}
The amplitude amplification will require an average number of calls to the circuit in figure \ref{fig:HHL_schematic} of order $\sqrt{1/\t p}$.
We are more restrictive on the state component to be amplified, $\t p<p_0$, but
the bound \eqref{eq:Gl-convergence} and $C=2\pi k_\nmin/t_0\simeq1/(2\kappa')$ prove that $\t p$ and $p_0$ are both of order $\Ord(\Gl_\nmin\,\kappa')^{-2}=\Ord(\kappa/\kappa')^2$.
Therefore, the HHL variant and the improved algorithm have the same scaling of the running time.

After the amplitude amplification of the state $\ket{1}_a\ket{0}_c$, the output state is
\begin{equation} \label{eq:out_c}
    \ket{\t x} = \frac{1}{\sqrt{\t p}} \sum_j \sum_{k=k_\nmin}^{T-1} \Gb_j \, \big|\Ga_{k|j}^\nvar\big|^2  \,\frac{Ct T}{2\pi k}\, \ket{1}_a\ket{0}_c\ket{u_j}_r \,.
\end{equation}
In analogy to the state \eqref{x.id} that
encodes the exact solution $\ve{x}$ in the Hilber space of quantum registers, the state \eqref{eq:out_c} encodes the approximate solution $\t{\ve{x}}=\sum_j \Gb_j \sum_{k=k_\nmin}^{T-1} \big|\Ga_{k|j}^\nvar\big|^2  \,\frac{t T}{2\pi k}\ve{u}_j$.
We then measure the error of the estimate through the fidelity between $\ket{\t x}$ and $\ket{x}$,
\begin{align}\label{eq:Fid_x_out_c}
    \braket{x}{\t x} 
    & = \frac{C^2}{\sqrt{p \t p}} \, \sum_{j=0}^{2^{n}-1} \frac{|\Gb_j|^2}{\Gl_j} \, \sum_{k=1}^{T-1} \big|\Ga_{k|j}^\nvar\big|^2\,\frac{tT}{2\pi k}
    \nonumber \\
    & = \frac{\displaystyle \sum_j \frac{|\Gb_j|^2}{\Gl_j} \sum_{k=1}^{T-1} \big|\Ga_{k|j}^\nvar\big|^2 \, \frac{t T}{2\pi k}}{\sqrt{\displaystyle \sum_j \frac{|\Gb_j|^2}{\Gl_j^2}
    \sum_l |\Gb_l|^2 \left(\sum_{k=1}^{T-1} \big|\Ga_{k|l}^\nvar\big|^2 \, \frac{t T}{2\pi k}\right)^2}} \,.
\end{align}
The difference between equation \eqref{eq:Fid_x_out_c} and equation \eqref{eq:Fid_x_out} is that the same sum appears in both the numerator and the denominator of \eqref{eq:Fid_x_out_c}. Then, the convergence of the distance $d\big(\ket{x},\ket{\t{x}}\big)$ depends only of the quantity $\Ge_{1,j}^\nvar$ \eqref{epsilon1} and on its bound \eqref{eq:Gl-convergence}. Using the notation $1/\t{\Gl}_j:=\sum_{k=k_\nmin}^{T-1} \big|\Ga_{k|j}^\nvar\big|^2 t T/(2\pi k)$, the computational error is given by
\begin{equation} \label{eq:dist.imp.var}
d\big(\ket{x},\ket{\t{x}}\big)=
\sqrt{
\frac{\displaystyle \sum_j \frac{|\Gb_j|^2}{\Gl_j^2} \sum_l \frac{|\Gb_l|^2}{\Gl_l\,\t{\Gl_l}} \left(
\frac{\Gl_l}{\t{\Gl_l}} - \frac{\Gl_j}{\t{\Gl_j}}  \right)} {\displaystyle \sum_j \frac{|\Gb_j|^2}{\Gl_j^2}
\sum_l \frac{|\Gb_l|^2}{\t{\Gl_l}^2}}
}\,.
\end{equation}
Since equation \eqref{eq:Gl-convergence} implies that $\Gl_j/\t{\Gl_j} = 1 + \Ord\left(\ln(\Gl_j \,t \,T)/(\Gl_j \,t \,T)\right)$, the error is bigger if the eigenvalue is smaller. Then, recalling $\Gl_\nmin = \min_j\Gl_j$, the error is bounded by
\begin{equation}\label{eq:error}
d\big(\ket{x},\ket{\t{x}}\big)
\leqslant
\Ord \left( \sqrt{\frac{\ln(\Gl_\nmin \,t\,T)}{\Gl_\nmin \,t\,T}} \right)
\leqslant
\Ord \left( \sqrt{\frac{\kappa}{t\,T}\,\ln\frac{t\,T}{\kappa}} \right)
\,.
\end{equation}
Equation \eqref{eq:error} shows that the error can be
arbitrary small by choosing a big enough number of qubits $n_c$ in the $c$ register, with $T=2^{n_c}$, provided that the condition number $\kappa$ is not as large as $T$.

The quantum states $\ket{\t x}$ and $\ket{x}$ have unit normalization, therefore, vectors $\t{\ve{x}}\in\C^d$ that differ only in the normalization have the same encoding $\ket{\t x}$. In this sense, the states $\ket{x}$ and $\ket{\t x}$ encode only the direction of the vectors $\ve{x}$ and $\t{\ve{x}}$ respectively. Nevertheless, the norms of the vectors $\t{\ve{x}}$ and $\ve{x}$ defined above are
\begin{equation}
    ||\t{\ve{x}}||_2 = \frac{\sqrt{\t p}}{C} \,,
    \qquad
    ||\ve{x}||_2 = \frac{\sqrt{p}}{C} \,.
\end{equation}
The probability $\t p$ can be measured implementing the amplitude estimation algorithm (a combination of the amplitude amplification and the quantum phase estimation algorithms) \cite{Boyer_1998,Brassard_2002} that uses an average number of calls to the circuit of figure \ref{fig:HHL_schematic} of order $1/\sqrt{\t p}$.
In order to prove that $||\t{\ve{x}}||_2$ well approximates $||\ve{x}||_2$, use equation \eqref{eq:Gl-convergence} and $\Gl_\nmin = \min_j \Gl_j$ to compute

\begin{equation} \label{Norm_convergences}
    \frac{\big|\,||\ve{x}||_2^2-||\t{\ve{x}}||_2^2\,\big|}{||\ve{x}||_2^2} \leqslant \frac{\displaystyle \sum_j \frac{|\Gb_j|^2}{\Gl_j^2} \left| 1 - \frac{\Gl_j^2}{\t{\Gl_j}^2} \right|}{\displaystyle \sum_j\frac{|\Gb_j|^2}{\Gl_j^2}} =\Ord\left(\frac{\ln(\Gl_\nmin \,t\,T)}{\Gl_\nmin \,t\,T} \right) \,.
\end{equation}
The bound \eqref{Norm_convergences} proves that the norm of the solution can be measure with
arbitrary small error
and with the same running time scaling required for the amplification of the amplitude of state $\ket{\t x}$.

The components of $\ve{x}$ cannot be computed efficiently using this algorithm, as discussed in section \ref{sec:HHL}. Nevertheless, the algorithm allows to efficiently compute expectation values $\ve{x}^TM\ve{x}$ if $M$ is a matrix that can be efficiently implemented on a quantum computer. In order to prove it, assume without loss of generality that $M$ is Hermitian with eigenvalues $\{m_j\}_{j=1,\dots,d}$ and eigenvectors
$\{\ve{m}_j\}_{j=1,\dots,d}$, otherwise apply the following argument to the Hermitian and the anti-Hermitian part of $M$ independently.
Perform then a projective measurement on the output of the algorithm, $\ket{\t x}$, with projectors
$\big\{\ket{m_j}\bra{m_j}\big\}_{j=0,\dots,T-1}$.
After $\mu$ of such measurements, the statistical frequency of obtaining the outcome $m_j$ approximates the probability $|\braket{\t x}{m_j}|^2$ up to an error $\delta$ with probability $1-\mathcal{O}(e^{-2\mu\delta^2})$ from the Chernoff-Hoeffding bound \cite{Chernoff1952,Hoeffding1963}. When the sampling error $\delta$ is comparable with the computational errors \eqref{eq:error},
one requires $\mu>\mathcal{O}\big(\Gl_\nmin\,t\,T/\ln(\Gl_\nmin \,t\,T)\big)$ repetitions for a good success probability in the Chernoff-Hoeffding bound. These repetitions can however be parallelized, without then increasing the computational time.

From the frequency of obtaining the state $\ket{m_j}$ and from the meaurement of the norm discussed above, one computes
\begin{equation} \label{eq:est.Mav}
\|\t{\ve{x}}\|_2^2\,
\sum_{j=0}^{T-1}m_j |\braket{\t x}{m_j}|^2
=\|\t{\ve{x}}\|_2^2\,\bra{\t x}M\ket{\t x}
=\t{\ve{x}}^TM\t{\ve{x}}
=\ve{x}^TM\ve{x}\left(1+\Ord\left(\frac{\ln(\Gl_\nmin \,t\,T)}{\Gl_\nmin \,t\,T}\right)\right)
 \,,
\end{equation}
achieving a good estimate of $\ve{x}^TM\ve{x}$ for large $T$.


\section{Complexity} \label{sec:compl}

We analyse and compare here the complexity of the above algorithms.
The relevant parameters for this analysis are
the computational error $\Ge$, the condition number $\Gk=\max_j|\Gl_j|/\min_j|\Gl_j|$ of the matrix $A$ and its sparsity $s$ that it the maximum number of non-vanishing elements in each row or column.

The simulation of the controlled gates
\eqref{eq:CexpA} within error $\epsilon'$ in the quantum phase estimation can be done with a number $\Ord\big(s\,T\max_{i,j}|A_{i,j}|+\frac{\ln(1/\epsilon')}{\ln\ln(1/\epsilon')}\big)$ of queries to the matrix elements of $A$ \cite{Low_2017,Low_2019} and additional $\Ord\big((\ln d)\,s\,T\max_{i,j}|A_{i,j}|+\frac{(\ln d)\ln(1/\epsilon')}{\ln\ln(1/\epsilon')}\big)$ one- and two-qubit gates.
Note that the assumption $0<\lambda_j\leqslant1$ implies $\max_{i,j}|A_{i,j}|\leqslant1$. The implementation of the quantum Fourier transform requires a number $\Ord(\log_2^2 T)$ of one- and two- qubit gates
\cite{nielsen2010quantum,BCRS2019} and the controlled rotations \eqref{eq:conditionated_Rotation} require $\Ord(T\log_2^2 T)$
of such operations (see appendix \ref{app:steps}).
Moreover, the circuit in figure \ref{fig:HHL_schematic} has to be executed $\Ord\big(1/\sqrt{\t p}\big) = \Ord(\kappa'/\kappa)$ times to implement the postselection.
The number of sequential queries is called \emph{query complexity}, $\mathcal{Q}=\Ord(s\,T\,\kappa'/\kappa)$,
and the number of consecutive additional gates is called \emph{gate complexity}, $\mathcal{G}=\Ord\big((s\log_2 d+\log_2^2T)\,T\,\kappa'/\kappa\big)$.
We focus on the gate complexity as the query complexity is proportional to $\mathcal{G}$ up to logarithmic corrections.

The gate complexity can be expressed in terms of the computational error quantified by the distance between the output state and the ideal state (see equations \eqref{eq:err.HHL} and \eqref{eq:dist.imp.var}).
From equation \eqref{eq:errHHL}, consider
the computational error $\Ge=\Gk/T$, and the gate complexity becomes $\mathcal{G}=\Ord\big((s\log_2 d+\log_2^2(\Gk/\Ge))\,\kappa'/\Ge\big)$.
Near term experimental realizations and applications of HHL algorithms as subroutines \cite{cao2012quantum,barz2013solving,Pan_2014,dervovic2018quantum,Duan2020,
morrell2023stepbystep}
could benefit from avoiding the preparation of the initial entangled state of the clock \eqref{eq:clock.HHL}. We then analyse the gate complexity of the HHL variant and of the improved HHL variant.

For the HHL variant, we consider the best computational error, namely the scaling of the lower bound \eqref{eq:error0}, while the computational error of the improved HHL variant is given by the upper bound \eqref{eq:error} which captures the exact numerical scaling.
The computational error is then $\Ge=\sqrt{\frac{\Gk}{T}\,\ln\left(\frac{T}{\Gk}\right)}$ for HHL variant with $\Gk'=\Ord\big(\sqrt{T\Gk\ln T/\Gk}\big)$ and for the improved HHL variant for any $\Gk'$. Inverting this relation, we obtain $T=-\frac{\Gk}{\Ge^2}\,W_{-1}(-\Ge^2)$ where $W_{-1}(z)$ is the branch of the Lambert function defined as the real solution of $We^W=z$ with largest absolute value $|W|$ \cite{hndbook2010}.

The gate complexity in this case becomes
\begin{equation} \label{gc.iv}
\mathcal{G}=\Ord\left(
-\frac{\Gk'}{\Ge^2}\,W_{-1}\big(-\Ge^2\big)
\left(s\log_2 d+\log_2^2\left(-\frac{\Gk'}{\Ge^2}\,W_{-1}\big(-\Ge^2\big)\right)\right)
\right) \,.
\end{equation}
Using the notation $\t\Ord(\cdot)$ which denotes orders $\Ord(\cdot)$ up to multiplicative powers of the logarithm of its argument, the gate complexity \eqref{gc.iv} reads $\mathcal{G}=\t{\Ord}(\Gk'\,s\,\Ge^{-2}\log_2d)$.
The condition $\Gk'=\Ord\big(\sqrt{T\Gk\ln T/\Gk}\big)$ for the HHL variant is compatible with $\Gk\leqslant\Gk'$ only if $\Gk\leqslant\chi\sqrt{T\Gk\ln T/\Gk}$ for some constant $\chi$. This inequality imples $\Gk\leqslant T\,e^{-W_0(1/\chi^2)}=T\,\chi^2\,W_0(1/\chi^2)$, where $W_0(z)$ is the principle branch of the Lambert function defined as the unique real solution of $We^W=z$ for positive $z$ (for instance $W_0(1)\simeq0.57$) \cite{hndbook2010}.

Equation
\eqref{eq:error0}
also imples that the computational error of the HHL variant with $\kappa'$ as large as $T$ ($\kappa'\sim T$) is a non-negligible constant, and the algorithm output is not reliable.
If, instead, $\Gk'\gg\sqrt{T\Gk\ln T/\Gk}$ then $\Gk\ll\t\Ord\big((\Gk')^2/T\big)$
and the gate complexity is lower bounded by $\mathcal{G}=\t\Ord\big(T\Gk'\Gk^{-1}s\,\log_2d\big)\gg\t\Ord\big(T^2(\Gk')^{-1}s\,\log_2d\big)$
where, given the computational error
$\Ge\sim\Gk'/T$ as in \eqref{eq:error0},
the bound equals the gate complexity $\t\Ord\big(\Gk'\,s\,\Ge^{-2}\log_2d\big)$ in \eqref{gc.iv}.
In conclusion,
for $\Gk'=\Ord\big(\sqrt{T\Gk\ln T/\Gk}\big)$ the HHL variant and the improved algorithm have comparable computational error, query and gate complexity, while
for $\Gk'\gg\sqrt{T\Gk\ln T/\Gk}$ the improved algorithm is more efficient in terms of these figures of merit.

\section{Numerical examples} \label{sec:num}

In this section, we simulate the HHL variant and the improved algorithm for the solution of two dimensional linear problems ($d=2$) using the python library qiskit, without implementing environmental noise on the quantum gates. We then need only one qubit in the $r$-register, and fix the parameters, $k_\nmin=1$, $t = 8\pi/5$ and $C= 2\pi/t_0$. The condition $k_\nmin=1$ corresponds to the worst case scenario without any bound of the condition number. The number of qubits $n_c$ of the $c$-register varies from $3$ to $11$.
For these numbers $n_c$, our classical simulation stores efficiently the full state vector, and then the amplitude amplifications is replaced by the readout of the corresponding state coefficient.
The code can be find on GitHub\footnote{https://github.com/MatiasGinzburg/Improvement-on-the-Convergence-in-Quantum-Linear-System-Solvers}.

\begin{figure}[h]
\includegraphics[width=0.5\textwidth]{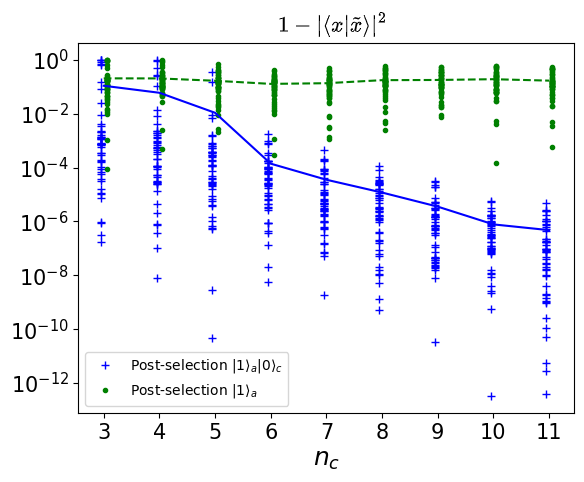}
\includegraphics[width=0.5\textwidth]{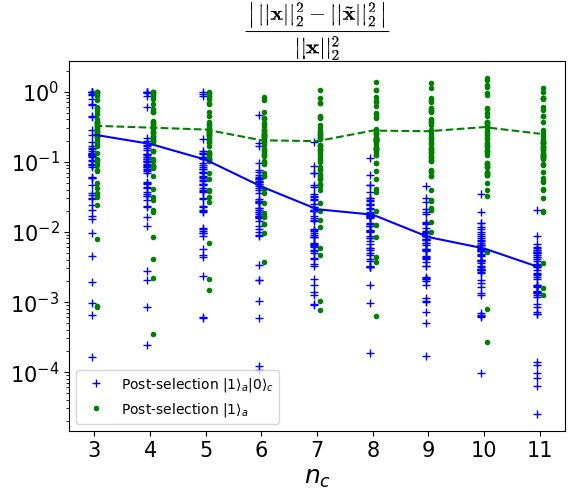}
\caption{Log-lin plot of the infidelity
(left panel) and of the norm error
(right panel) between the output state and the ideal state of 50 randomly generated linear problems.
Green dots corresponds to the standard HHL algorithm with postselection of the state $|1\rangle_a$, whose average behaviours are the dashed green lines. Blue crosses are the errors of the improved algorithm, and the solid blue lines are the averages.}
\label{fig:num}
\end{figure}

We randomly generated 50 linear problems of the form \eqref{eq:Ax=b} defined by Hermitian matrices $A$ with eigenvalues in the range $(0,1]$
and vectors $b$ with unit norm. To generate such random matrices $A$, we sampled diagonal matrices $A'=\begin{pmatrix}\lambda_1 & 0 \\ 0 & \lambda_2\end{pmatrix}$
from the uniform distribution of $\lambda_j\in(0,1]$
and transform them through a random unitary matrix, $A=UA'\,U^{\dag}$, with $U$ drawn from the Haar measure \cite{mezzadri2006generate}. In figure \ref{fig:num}, we plot the infidelity
(left panel) and the norm error
(right panel) as function of the number $n_c$ of qubits of the clock.
We compared the HHL variant with postselection of the state $|1\rangle_a$ (green dots) and our improved algorithm with postselection of the state $|1\rangle_a|0\rangle_c$ (blue crosses), and the lines represent the average behaviours. The plots show that the errors of the HHL variant remain constant, while the errors of the new algorithm decrease to zero with increasing $n_c$ as computed in section \ref{sec:new}.

\section{Negative eigenvalues of $A$} \label{sec:neg}

In this section, we discuss how to apply the proposed algorithm to the case of non-positive eigenvalues of $A$. Using the normalization argument described in section \ref{sec:HHL}, a general linear problem \eqref{eq:Ax=b} is transformed into a linear problem with $-1\leqslant\Gl_j\leqslant1$.

First, set $t_0=tT$ as before but with $0<t<\pi$.
After the phase estimation algorithm the distribution $|\Ga_{k|j}|^2$ is centered around $k\sim \Gl_j tT/(2\pi) \text{ mod } T $.
When the eigenvalue is positive $\lambda_j>0$, the most probable value is $k\sim \Gl_j tT/(2\pi)<T/2$. When the eigenvalue is negative $\lambda_j<0$, the most probable value is $k\sim \Gl_j tT/(2\pi) \text{ mod } T > T/2 $. In other words, the periodicity of the $\Ga_{k|j}$ is exploited to differentiate between positive and negative eigenvalues. Next, the angles $\theta(k)$ of the conditional rotation \eqref{eq:conditionated_Rotation} are defined by
\begin{equation} \label{eq:theta(k)negativelambda}
\sin\big(\theta(k)\big) :=
\begin{cases}
\displaystyle \frac{C \, t_0}{2\pi k}
& \displaystyle \textnormal{if } k\leqslant T/2 \\
\displaystyle \frac{C \, t_0}{2\pi(k-T)}
& \displaystyle \textnormal{if } k\geqslant T/2
\end{cases}
\,.
\end{equation}

The bounds
\eqref{eq:Gl-convergence} and \eqref{eq:Gl_square} hold true
also for this case and so do all the conclusions. The difference in the computations arises in equation \eqref{eq:sine_Bound}, where we should use the new equation \eqref{eq:theta(k)negativelambda}. We also have to consider the cases $0<t\,\Gl_j/(2\pi)<1/2$ and $-1/2<t\,\Gl_j/(2\pi)<0$ separately in appendix \ref{app:bound1}. There are divergences for $|t\,\Gl/(2\pi)|\approx 1/2$ which are avoided by choosing $t\neq\pi$. With this little modification, more general matrices can be handled.

\section{Conclusions} \label{sec:concl}

We have analysed the HHL algorithm and a commonly used variant of the HHL algorithm for solving linear systems \eqref{eq:Ax=b} on quantum computers.
Both these algorithms consist in two main subroutines, the quantum phase estimation and the amplitude amplification.
The quantum phase estimation is used for computing the eigenvalue of the matrix $A$ in \eqref{eq:Ax=b}, and the amplitude amplification is required for increasing the success probability of the matrix inversion.
The HHL variant replaces the initial entangled state \eqref{eq:clock.HHL} of the auxiliary register used in the quantum phase estimation, called clock, with the fully factorizable state \eqref{eq:clock.var}.

We analysed how the performances of the HHL algorithm and of its variant scale with the number of qubits $n_c$ of the clock. This parameter is relevant as it gives the precision of the quantum phase estimation for eigenvalue computations and the number of required gates.
We proved that, the error of the HHL variant, unlike that of the original HHL algorithm, does not converge for large $n_c$ but instead oscillates around a constant value,
when the condition number of the matrix $A$ is large or when a poor estimation of it is used.
The reason is the presence of oscillating terms in the amplified amplitude. To solve this problem we proposed a modified algorithm, based on the amplification of a smaller amplitude without detrimental oscillating terms.
We confirmed this result with numerical simulations.

The simulation of unitary exponentials of $A$ in the quantum phase estimation contribute significantly to the gate counting in previous discussions of the original HHL algorithm and of its variant.
We have therefore studied the complexity of these algorithms in the light of recent results on unitary similations
\cite{berry2007efficient,BerryChilds2012,
BerryCleveGharibian2014,Berry2014,Berry_2015,Berry2015-2,BerryNovo2016,BerryNovo2017,
Low_2017,Low_2019}.
Our improved algorithm has a lower complexity than the HHL variant
not only when the prior estimation of the condition number is so large that the error of the HHL variant does not converge, but also at smaller condition numbers.

Our results are relevant for applications of the HHL algorithm and its variants as subroutines of more complicated algorithms and for next term experimental realisations
with noisy intermediate-scale quantum (NISQ) technologies \cite{Preskill2018,Bharti2022}.

\appendix

\section{Steps of the HHL variant} \label{app:steps}

\begin{figure}[hb]
    \centering
    \includegraphics[width=0.4\textwidth]{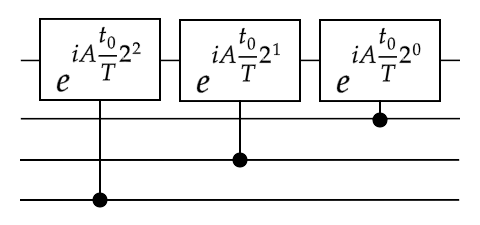}
    \caption{Implementation of the conditional evolution operation for a clock with 3 qubits.}
    \label{fig:conditionalExponential}
\end{figure}

In this appendix, we explicitly show intermediate states of the circuit in figure \ref{fig:HHL_schematic}.
The first operator $\mathcal{C}$ is the initialization of the clock register that differentiates the original HHL algorithm from its variant, as discussed in section \eqref{sec:HHL}.
The second gate \eqref{eq:CexpA}
can be realized using the circuit represented in figure \ref{fig:conditionalExponential} \cite{Cleve_1998} where each of the exponentials can be implemented if the matrix $A$ is sparse
or if it is a linear combination of unitaries
\cite{berry2007efficient,BerryChilds2012,
BerryCleveGharibian2014,Berry2014,Berry_2015,Berry2015-2,BerryNovo2016,BerryNovo2017,
Low_2017,Low_2019}.
The state after these operations is
\begin{align}
\ket{\psi_1^\nhhl} & =\ket{0}_a \sqrt{\frac{2}{T}} \, \sum_{j=0}^{2^n-1} \sum_{\Gt=0}^{T-1} \Gb_j \, \sin\left(\frac{\pi\left(2\,\tau+1\right)}{2\,T}\right) \, \e{i\frac{\Gt}{T}\Gl_j t_0} \ket{\Gt}_c \ket{u_j}_r \,, \label{eq:psi_2HHL} \\
\ket{\psi_1^\nvar} & =\ket{0}_a \frac{1}{\sqrt{T}} \, \sum_{j=0}^{2^n-1} \sum_{\Gt=0}^{T-1} \Gb_j \, \e{i\frac{\Gt}{T}\Gl_j t_0} \ket{\Gt}_c \ket{u_j}_r \,, \label{eq:psi_2}
\end{align}
for the orginal HHL algorithm and for the HHL variant respectively.

After the inverse quantum Fourier transform the state vector can be written in the following compact form for both the HHL algorithm and the HHL variant,
\begin{equation}\label{eq:psi_3}
    \ket{\psi_2} = \ket{0}_a \sum_{j=0}^{2^n-1} \sum_{k=0}^{T-1} \Gb_j \, \Ga_{k|j} \, \ket{k}_c \ket{u_j}_r \,,
\end{equation}
where the coefficients $\Ga_{k|j}$ are defined in equations \eqref{eq:alpha.HHL} and \eqref{eq:alpha}.
Since the state $\ket{\psi_2}$ has been achieved through unitary operations, it is normalized such that
\begin{equation} \label{eq:norm.alpha}
1=\langle \psi_2|\psi_2\rangle=\sum_{k=0}^{T-1}\big|\Ga_{k|j}\big|^2 \,.
\end{equation}

\begin{figure}
    \centering
    \includegraphics[width=0.5\textwidth]{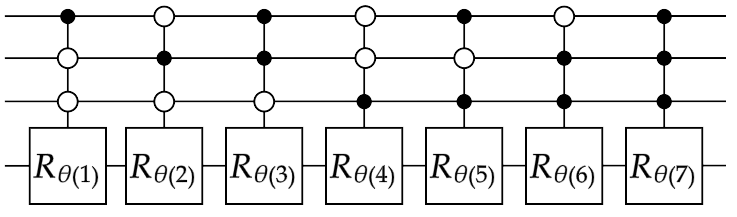}
    \caption{Implementation of the conditional rotation describe in equation \eqref{eq:conditionated_Rotation} with $k_\nmin=1$ and $n_c=3$ qubits of the clock.}
    \label{fig:conditionalRotation}
\end{figure}

The next gate, namely the controlled unitary defined by the transformation \eqref{eq:conditionated_Rotation}, can be implemented using $2^{n_c}-k_\nmin$ multi-qubit controlled rotations, as exemplified in figure \ref{fig:conditionalRotation}. Each multi-qubit controlled rotation requires $\mathcal{O}(n_c^2)$ elementary gates \cite{Barenco1995}, and thus the transformation \eqref{eq:conditionated_Rotation} is implemented in time $\mathcal{O}(n_c^2\,2^{n_c})$.
The state vector after this step is
\begin{align}\label{eq:psi_4}
\ket{\psi_3} =& \sum_{j=0}^{2^{n}-1} \sum_{k=k_\nmin}^{T-1} \Gb_j \, \Ga_{k|j} \, \sin\big(\theta(k)\big) \ket{1}_a\ket{k}_c \ket{u_j}_r \nonumber \\
    & + \sum_{j=0}^{2^{n}-1} \sum_{k=k_\nmin}^{T-1} \Gb_j \, \Ga_{k|j} \, \cos\big(\theta(k)\big) \ket{0}_a\ket{k}_c\ket{u_j}_r  \nonumber \\
    & + \sum_{j=0}^{2^{n}-1} \sum_{k=0}^{k_\nmin-1} \Gb_j \, \Ga_{k|j} \, \ket{0}_a\ket{0}_c\ket{u_j}_r\,,
\end{align}
recalling $\sin{\theta(k)} := C t_0/(2\pi k)$.

The quantum Fourier transform then yields the state vector
\begin{align}\label{eq:psi_5}
    \ket{\psi_4} = & \frac{1}{\sqrt{T}} \sum_{j=0}^{2^{n}-1} \sum_{\Gs=0}^{T-1} \sum_{k=k_\nmin}^{T-1} \Gb_j \, \Ga_{k|j} \, \sin\big(\theta(k)\big) \, \e{i \frac{\Gs}{T}2\pi k} \ket{1}_a\ket{\Gs}_c\ket{u_j}_r \nonumber \\
    & + \frac{1}{\sqrt{T}} \sum_{j=0}^{2^{n}-1} \sum_{\Gs=0}^{T-1} \sum_{k=k_\nmin}^{T-1} \Gb_j \, \Ga_{k|j} \, \cos\big(\theta(k)\big) \, \e{i \frac{\Gs}{T}2\pi k} \ket{0}_a\ket{\Gs}_c\ket{u_j}_r \nonumber \\
    & + \frac{1}{\sqrt{T}} \sum_{j=0}^{2^{n}-1} \sum_{\Gs=0}^{T-1} \sum_{k=0}^{k_\nmin-1} \Gb_j \, \Ga_{k|j} \, \ket{0}_a\ket{\Gs}_c\ket{u_j}_r \,.
\end{align}

After the inverse of the transformation \eqref{eq:CexpA} and the last gate $\mathcal{C}^{-1}$, one obtains the state \eqref{eq:psi_7}.

\section{Computation of the bound \eqref{eq:Gl-convergence}} \label{app:bound1}

In this appendix we derive the bound \eqref{eq:Gl-convergence}.
First, use the identity \eqref{eq:norm.alpha} and the triangle inequality to compute
\begin{align}
|\Ge_{1,j}| &= \Gl_j \left| \frac{1}{\Gl_j}-\sum_{k=k_\nmin}^{T-1} \big|\Ga_{k|j}^\nvar\big|^2  \frac{t T}{2\pi k} \right| \nonumber \\
& =\left|\sum_{k=0}^{k_\nmin-1} \big|\Ga_{k|j}^\nvar\big|^2 + \frac{\sin^2\left(\frac{\Gl_j t T}{2}\right)}{T^2} \sum_{k=k_\nmin}^{T-1} \frac{\left(k-\frac{\Gl_j t T}{2\pi}\right)}{k\sin^2\left(\frac{\pi}{T}\left(k-\frac{\Gl_j t T}{2\pi}\right)\right)}\right| \nonumber \\
& \leqslant \sum_{k=0}^{k_\nmin-1} \big|\Ga_{k|j}^\nvar\big|^2 + \frac{\sin^2\left(\frac{\Gl_j t T}{2}\right)}{T^2} \sum_{k=k_\nmin}^{T-1} \frac{\left|k-\frac{\Gl_j t T}{2\pi}\right|}{k\sin^2\left(\frac{\pi}{T}\left(k-\frac{\Gl_j t T}{2\pi}\right)\right)} \,.
\label{eq:sine_Bound}
\end{align}
In order to bound the sums in the right-hand-side of \eqref{eq:sine_Bound}, we will extensively use the following inequalities
\begin{align}
\label{eq:sin.lb}
\sin(\pi x)
& \geqslant
\begin{cases}
2x & \text{if } 0 \leqslant x \leqslant\frac{1}{2} \\
2(1-x) & \text{if } \frac{1}{2}\leqslant x \leqslant 1
\end{cases}\,, \\
\label{eq:sum.ub}
\sum_{x=a}^{b} f(x)
& \leqslant \int _a^b f(x) dx + f(a) + f(b) \,,
\end{align}
valid for positive convex functions $f(x)$.
The two cases in inequality \eqref{eq:sin.lb} require to separatedly treat different contributions in the upper bound \eqref{eq:sine_Bound}.

Let us start with considering $t\Gl_j/(2\pi)\leqslant 1/2$. The first term in \eqref{eq:sine_Bound} is bounded as follows:
\begin{align}\label{eq:boundSoloAlpha}
& \sum_{k=0}^{k_\nmin-1} \big|\Ga_{k|j}^\nvar\big|^2
=
\sum_{k=0}^{k_\nmin-1} \frac{\sin^2\left(\frac{\Gl_j t T}{2}\right)}{T^2\,\sin^2\left(\frac{\pi}{T}\left(\frac{\Gl_jtT}{2\pi}-k\right)\right)}
\leqslant \frac{1}{4} \, \sin^2\left(\frac{\Gl_j t T}{2}\right) \sum_{k=0}^{k_\nmin-1} \left(\frac{\Gl_jtT}{2\pi}-k\right)^{-2} \nonumber \\
& \leqslant
\frac{\sin^2\left(\frac{\Gl_j t T}{2}\right)}{4} \left( \int_{k=0}^{k_\nmin-1} \left(\frac{\Gl_jtT}{2\pi}-k\right)^{-2}\, dk
+ \left(\frac{2\pi}{\Gl_jtT}\right)^2 + \left(\frac{\Gl_jtT}{2\pi}-k_\nmin+1\right)^{-2} \right) \nonumber\\
& \leqslant
\frac{\sin^2\left(\frac{\Gl_j t T}{2}\right)}{4} \left( \frac{k_\nmin-1}{\frac{\Gl_jtT}{2\pi}\left(\frac{\Gl_jtT}{2\pi}-k_\nmin+1\right)} + \left(\frac{2\pi}{\Gl_jtT}\right)^2 + \left(\frac{\Gl_jtT}{2\pi}-k_\nmin+1\right)^{-2}  \right) \nonumber \\
& \leqslant \frac{\pi}{2\Gl_j tT} + \left( \frac{2\pi}{\Gl_j t T}\right)^2 = \Ord\left( \frac{1}{\Gl_j t T} \right) \,,
\end{align}
where we have used the assumption $1\leqslant k_\nmin<\Gl_jtT/(4\pi)$ (see section \ref{sec:HHL}) in the last inequality.

Defining $a_j = \lfloor \Gl_j t T /2\pi \rfloor$ and $0\leqslant \GD_j< 1$ such that
\begin{equation} \label{def:a-delta}
    \frac{\Gl_j t T}{2\pi} = a_j + \GD_j \,,
\end{equation}
the second sum in the upper bound \eqref{eq:sine_Bound} can be split into the following three sums
\begin{equation} \label{sum.split1}
\sum_{k=k_\nmin}^{T-1} = \sum_{k=k_\nmin}^{a_j} + \sum_{k=a_j+1}^{a_j+\frac{T}{2}} + \sum_{k=a_j+\frac{T}{2}+1}^{T-1} \,.
\end{equation}
Using again the inequalities \eqref{eq:sin.lb} and \eqref{eq:sum.ub},
we compute
for the first of these sums
\begin{align}
&\frac{\sin^2\left(\frac{\Gl_j t T}{2}\right)}{T^2}\sum_{k=k_\nmin}^{a_j} \frac{\frac{\Gl_j t T}{2\pi}-k}{k \, \sin^2\left(\frac{\pi}{T}\left(\frac{\Gl_j t T}{2\pi}-k\right)\right)} \nonumber \\
& \leqslant \frac{\sin^2\left(\frac{\Gl_j t T}{2}\right)}{4} \sum_{k=k_\nmin}^{a_j} \frac{1}{k\,\left(\frac{\Gl_j t T}{2\pi}-k\right)} \nonumber \\
&\leqslant \frac{\sin^2\left(\frac{\Gl_j t T}{2}\right)}{4}\left(\int_{k_\nmin}^{a_j}  \frac{1}{k\,\left(\frac{\Gl_j t T}{2\pi}-k\right)} \,dk + \frac{1}{k_\nmin\left(\frac{\Gl_j t T}{2\pi}-k_\nmin\right)} +  \frac{1}{\GD_j\,a_j} \right) \nonumber \\ 
&= \frac{\sin^2\left(\frac{\Gl_j t T}{2}\right)}{4}\left( \frac{2\pi}{\Gl_j t T} \ln\left(\frac{a_j\,\left(\frac{\Gl_j t T}{2\pi}-k_\nmin\right)}{\GD_j\,k_\nmin } \right) + \frac{1}{k_\nmin\left(\frac{\Gl_j t T}{2\pi}-k_\nmin\right)} + \frac{1}{\GD_j\,a_j} \right) \nonumber \\
& \leqslant \frac{\pi}{\Gl_j t T} \ln \left(\frac{\Gl_j t T}{2\pi}\right) + \frac{12\pi}{4\Gl_j t T} = \Ord\left(\frac{\ln(\Gl_j t T)}{\Gl_j t T} \right) \,,
\label{est.sum1.1}
\end{align}
where we have used, in the last inequality, $\sin^2(\Gl_j t T/2)=\sin^2(\pi\GD_j)$, $0<-\sin^2(\pi\GD_j)\ln(\GD_j)<1$ and $\frac{\sin^2(\pi\GD_j)}{\GD_j} < 5/2$.

A very similar calculation bounds the second sum in \eqref{sum.split1}:
\begin{align}
&\frac{\sin^2\left(\frac{\Gl_j t T}{2}\right)}{\,T^2} \sum_{k=a_j+1}^{\frac{T}{2}+a_j} \frac{k-\frac{\Gl_j t T}{2\pi}}{k\,\sin^2\left(\frac{\pi}{T}\left(k-\frac{\Gl_j t T}{2\pi}\right)\right)} \nonumber \\
& \leqslant \frac{\sin^2\left(\frac{\Gl_j t T}{2}\right)}{4} \sum_{k=a_j+1}^{\frac{T}{2}+a_j} \frac{1}{k\,\left(k-\frac{\Gl_j t T}{2\pi}\right)} \nonumber \\
& \leqslant \frac{\sin^2\left(\frac{\Gl_j t T}{2}\right)}{4} \left( \frac{4}{\left(T-2\GD_j\right)\left(T+2a_j\right)} + \frac{1}{\left(1-\GD_j\right)\left(a_j+1\right)} + \int_{a_j+1}^{\frac{T}{2}+a_j} \frac{1}{k\,\left(k-\frac{\Gl_j t T}{2\pi}\right)}\,dk  \right) \nonumber \\
& =  \frac{\sin^2\left(\frac{\Gl_j t T}{2}\right)}{4} \left(\frac{4}{\left(T-2\GD_j\right)\left(T+2a_j\right)} + \frac{1}{\left(1-\GD_j\right)\left(a_j+1\right)} +  \frac{2\pi}{\Gl_j t T} \ln\frac{\left(T-2\GD_j\right)\left(a_j+1\right)}{\left(T+2a_j\big)(1-\GD_j\right)} \right) \nonumber \\
&\leqslant \frac{1}{4} \left( \frac{4}{T(T-2)} + \frac{10\pi}{2\Gl_j t T} -\frac{2\pi\ln 4}{\Gl_j t T} + \frac{2\pi}{\Gl_j t T} + \frac{2\pi}{\Gl_j t T}\ln\left(\frac{\Gl_j t T}{2\pi}+1\right)\right) \nonumber \\
&= \Ord\left(\frac{\ln(\Gl_j t T)}{\Gl_j t T} \right).
\label{est.sum1.2}
\end{align}

The last inequality in equation \eqref{est.sum1.2} follows from
the bounds $\frac{\sin^2(\pi\GD_j)}{1-\GD_j} < 5/2$, $-\sin^2(\pi\GD_j)\ln(1-\GD_j)<1$, and $ 1/2- 1/T < \big(T-2\GD_j\big)/\big(T+2a_j\big) < 1$ which comes from $\Gl_j t/(2\pi)\leqslant1/2$ and implies $\big|\ln\big(T-2\GD_j\big)/\big(T+2a_j\big)\big| < \ln 4$ for $T\geqslant4$.

Repeating the same procedure, the contribution from the last sum in \eqref{sum.split1} is
\begin{align}
&\frac{\sin^2\left(\frac{\Gl_j t T}{2}\right)}{\,T^2} \sum_{k=\frac{T}{2}+a_j+1}^{T-1} \frac{k-\frac{\Gl_j t T}{2\pi}}{k\,\sin^2\left(\frac{\pi}{T}\left(k-\frac{\Gl_j t T}{2\pi}\right)\right)} \nonumber \\
& \leqslant \frac{\sin^2\left(\frac{\Gl_j t T}{2}\right)}{4} \sum_{k=\frac{T}{2}+a_j+1}^{T-1} \frac{k-\frac{\Gl_j t T}{2\pi}}{k\,\left(T +\frac{\Gl_j t T}{2\pi}-k\right)^2} \nonumber \\
& \leqslant \frac{\sin^2\left(\frac{\Gl_j t T}{2}\right)}{4} \left(\rule{0cm}{1cm}\right. \frac{T-1-\frac{\Gl_j t T}{2\pi}}{(T-1)\left(\frac{\Gl_j t T}{2\pi}+1\right)^2} + \frac{4\left(T+2-2\GD_j\right)}{\left(T+ 2\GD_j-2\right)^2\,\left(T+2a_j+2\right)}  \nonumber \\
&+ \int_{\frac{T}{2}+a_j+1}^{T-1} \frac{k-\frac{\Gl_j t T}{2\pi}}{k\,\left(T + \frac{\Gl_j t T}{2\pi}-k\right)^2}\,dk \left.\rule{0cm}{1cm}\right) \nonumber \\
& = \frac{\sin^2\left(\frac{\Gl_j t T}{2}\right)}{4} \left(\rule{0cm}{1cm}\right. \frac{T-1-\frac{\Gl_j t T}{2\pi}}{(T-1)\left(\frac{\Gl_j t T}{2\pi}+1\right)^2} +\frac{4\left(T+2-2\GD_j\right)}{\left(T+ 2\GD_j-2\right)^2\,\left(T+2a_j+2\right)}  \nonumber \\ 
& + \frac{T}{T+\frac{\Gl_j t T}{2\pi}} \left( \frac{1}{\frac{\Gl_j t T}{2\pi}+1} - \frac{2}{T+2\GD_j-2} \right)  + \frac{\frac{\Gl_j t T}{2\pi}}{\left(T+\frac{\Gl_j t T}{2\pi}\right)^2} \ln\frac{\left(\frac{\Gl_j t T}{2\pi}+1\right)\left(T+2a_j+2\right)}{(T-1)\left(T+ 2\GD_j -2\right)} \left.\rule{0cm}{1cm}\right) \nonumber \\ 
& \leqslant \frac{1}{4} \left( \left(\frac{2\pi}{\Gl_j t T}\right)^2 + \frac{4}{T^2} + \frac{2\pi}{\Gl_j t T} + \frac{2\pi}{\Gl_j t T}\ln\left(2\left(\frac{\Gl_j t T}{2\pi}+1\right)\frac{T+1}{(T-1)^2}\right)\right) \nonumber\\
&= \Ord\left(\frac{1}{\Gl_j t T} \right)\,.
\label{est.sum1.3}
\end{align}
Finally, the estimates \eqref{eq:boundSoloAlpha}, \eqref{est.sum1.1}, \eqref{est.sum1.2}, and \eqref{est.sum1.3} imply \eqref{eq:Gl-convergence}  for $0<t\Gl_j/(2\pi)\leqslant 1/2$.

The second sum in the right-hand-side of \eqref{eq:sine_Bound} for the remaining case $1/2 < t\Gl_j/(2\pi) < 1$ is bounded following the same steps.
Split the sum as
\begin{equation} \label{sum.split2}
    \sum_{k=k_\nmin}^{T-1} = \sum_{k=k_\nmin}^{a_j-\frac{T}{2}} + \sum_{k=a_j-\frac{T}{2}+1}^{a_j} + \sum_{k=a_j+1}^{T-1} \,,
\end{equation}
and use the inequalities \eqref{eq:sin.lb} and \eqref{eq:sum.ub} appropriately.
The bound of the first sum in the right-hand-side of \eqref{sum.split2} is
\begin{align}
    &\frac{\sin^2\left(\frac{\Gl_j t T}{2}\right)}{T^2} \sum_{k=k_\nmin}^{a_j-\frac{T}{2}} \frac{\frac{\Gl_j t T}{2\pi}-k}{k \sin^2\left(\frac{\pi}{T}\left(\frac{\Gl_j t T}{2\pi}-k\right)\right)}
\leq \frac{\sin^2\left(\frac{\Gl_j t T}{2}\right)}{4} \sum_{k=k_\nmin}^{a_j-\frac{T}{2}} \frac{\frac{\Gl_j t T}{2\pi}-k}{k\,\left(T+k-\frac{\Gl_j t T }{2\pi}\right)^2} \nonumber \\
    &\leq \frac{\sin^2\left(\frac{\Gl_j t T}{2}\right)}{4}\left( \frac{\frac{\Gl_j t T}{2\pi}-k_\nmin}{k_\nmin \left(T+k_\nmin-\frac{\Gl_j t T }{2\pi}\right)^2} + \frac{4(T+2\GD_j)}{\left(T-2\GD_j\right)^2 \left(2a_j- T \right)} \right. \nonumber \\
    & \left. + \int_{k=k_\nmin}^{a_j-\frac{T}{2}} \frac{\frac{\Gl_j t T}{2\pi}-k}{k\, \left(T+k-\frac{\Gl_j t T }{2\pi}\right)^2}\,dk \right)  \nonumber \\
    & = \frac{\sin^2\left(\frac{\Gl_j t T}{2}\right)}{4} \scalebox{1.3}{\Bigg(}  \frac{\frac{\Gl_j t T}{2\pi}-k_\nmin}{k_\nmin\left(T+k_\nmin-\frac{\Gl_j t T }{2\pi}\right)^2} + \frac{4(T+2\GD_j)}{\left(T-2\GD_j\right)^2 \left(2a_j- T \right)} \nonumber \\
    & + \frac{T}{T-\frac{\Gl_j t T }{2\pi}} \left( \frac{2}{T-2\GD_j}- \frac{1}{T-\frac{\Gl_j t T }{2\pi}+k_\nmin} \right) \nonumber \\
    & + \frac{\frac{\Gl_j t T}{2\pi}}{\left(T-\frac{\Gl_j t T }{2\pi}\right)^2} \, \ln\left(\frac{\left(2a_j-T\right)\left(T-\frac{\Gl_j t T }{2\pi}+k_\nmin\right)}{k_\nmin\left(T-2\GD_j\right)} \right) \scalebox{1.3}{\Bigg)} \nonumber \\
    & = \Ord\left(\frac{\ln T}{T}\right) \,.
\end{align}
This bound shows terms that diverge when $\Gl_j t/(2\pi)$ approaches $1$, but the divergence is avoided by choosing $\max_j \Gl_j < 1$ and $t< 2\pi$. Indeed, $t$ is a free parameter of the algorithm, and $\max_j \Gl_j < 1$ can be achieved by a rescaling of the linear problem as discussed in Section \ref{sec:HHL}.

The other two sums in the right-hand-side of \eqref{sum.split2} are bounded following very similar computations. Finally, we obtain equation \eqref{eq:Gl-convergence} for any value $0 < \Gl_j t /(2\pi) <1$.

\section{Computation of the bound \eqref{eq:Gl_square}} \label{app:bound2}

In this appendix we compute the upper and lower bounds in equation \eqref{eq:Gl_square}. These bounds show the dependence of the computational error on the algorithm parameters $T$ and $k_\nmin$.
In particular, the error does not decrease with $T$ when $k_\nmin= \Ord(T^0)$ which corresponds to a lower bound of the conditional number $\Gk'=tT/(2\pi k_\nmin)=\Ord(T)$.

We start with upper bounding the error contribution \eqref{epsilon2} using the triangle inequality,
\begin{align} \label{rewr15}
|\Ge_{2,j}| &= \Gl_j^2 \left|\sum_{k=k_\nmin}^{T-1} \big|\Ga_{k|j}^\nvar\big|^2  \left(\frac{t T}{2\pi k}\right)^2 - \frac{1}{\Gl_j^2}\right| \nonumber \\
& \leqslant \sum_{k=0}^{k_\nmin-1} \big|\Ga_{k|j}^\nvar\big|^2 + \frac{\sin^2\left(\frac{\Gl_j tT}{2}\right)}{T^2} \sum_{k=k_\nmin}^{T-1} \frac{\left|\frac{\Gl_j t T}{2\pi}-k \right|\,\left(\frac{\Gl_j t T}{2\pi}+k\right)}{k^2\,\sin^2\left(\frac{\pi}{T}\left(\frac{\Gl_j t T}{2\pi}-k\right)\right)}
\,.
\end{align}
The first sum is bounded by $\Ord(\Gl_jtT)^{-1}$, as shown in equation \eqref{eq:boundSoloAlpha}. In order to bound the second sum, we follow the same steps as in the previous appendix. For concreteness, we show only the case $\Gl_j t/(2\pi) < 1/2$, because the other case is very similar. Splitting the sum as in equation \eqref{sum.split1} and recalling the notation \eqref{def:a-delta},
the first contribution is the largest one and is bounded as follows
\begin{align}
    & \frac{\sin^2\left(\frac{\Gl_j t T}{2}\right)}{T^2} \sum_{k=k_\nmin}^{a_j} \frac{\left|\frac{\Gl_j t T}{2\pi}-k \right| \left(\frac{\Gl_j t T}{2\pi}+k\right)}{k^2\,\sin^2\left(\frac{\pi}{T}\left(\frac{\Gl_j t T}{2\pi}-k\right)\right)} \leqslant \frac{\sin^2\left(\frac{\Gl_j t T}{2}\right)}{4} \sum_{k=k_\nmin}^{a_j} \frac{k+\frac{\Gl_j t T}{2\pi}}{k^2\,\left(\frac{\Gl_j t T}{2\pi}-k \right)} \nonumber \\
    &\leqslant \frac{\sin^2\left(\frac{\Gl_j t T}{2}\right)}{4} \left(  \int_{k_\nmin}^{a_j} \frac{k+\frac{\Gl_j t T}{2\pi}}{k^2\,\left(\frac{\Gl_j t T}{2\pi}-k \right)}\,dk + \frac{a_j + \frac{\Gl_j t T}{2\pi} }{\GD_j\,a_j^2} + \frac{\frac{\Gl_j t T}{2\pi}+ k_\nmin}{k_\nmin^2\left(\frac{\Gl_j t T}{2\pi}-k_\nmin\right)} \right) \nonumber \\
    & = \frac{\sin^2\left(\frac{\Gl_j t T}{2}\right)}{4} \Bigg( \frac{4\pi}{\Gl_j t T} \ln\left(\frac{a_j \left(\frac{\Gl_j t T}{2\pi}-k_\nmin\right)}{k_\nmin \GD_j}\right) \nonumber \\
    & + \frac{a_j-k_\nmin}{k_\nmin a_j } + \frac{2a_j + \GD_j}{\GD_j\,a_j^2} + \frac{\frac{\Gl_j t T}{2\pi}+ k_\nmin}{k_\nmin^2\left(\frac{\Gl_j t T}{2\pi}-k_\nmin\right)} \Bigg) \nonumber \\
    & \leqslant \Ord\left( \frac{\ln\left(\Gl_j t T\right)}{\Gl_j t T} \right) + \Ord\left(\frac{\kappa'}{tT} \right) \,,
    \label{eq:Gl_square_upper}
\end{align}
where we have used $(a_j-k_\nmin)/(k_\nmin(a_j))=\Ord\left({\kappa'}/{tT} \right)$ in the last inequality. Using once again the same techniques, all the other terms coming from the splitting \eqref{sum.split1} are bounded by orders $\Ord(\ln(\Gl_j t T)/(\Gl_j t T))$. We then prove the upper bound in equation \eqref{eq:Gl_square}.

For the computation of the lower bound, we look again at the sum with $k\in[k_\nmin,a_j]$.
Since all the terms are positive, we get a lower bound by neglecting all terms except $k=k_\nmin$,
\begin{align}\label{eq:Gl_square_ineq}
&\frac{\sin^2\left(\frac{\Gl_j t T}{2}\right)}{T^2}  \sum_{k=k_\nmin}^{a_j} \frac{\left(\frac{\Gl_j t T}{2\pi}-k\right)\left(\frac{\Gl_j t T}{2\pi}+k\right)}{k^2 \, \sin^2\left(\frac{\pi}{T}\left(\frac{\Gl_j t T}{2\pi}-k\right)\right)} \nonumber \\ 
& \geqslant \frac{\sin^2\left(\frac{\Gl_j t T}{2}\right)}{k_\nmin^2 T^2 } \cdot \frac{\left(\frac{\Gl_j t T}{2\pi}-k_\nmin\right)\left(\frac{\Gl_j t T}{2\pi}+k_\nmin\right)}{ \sin^2\left(\frac{\pi}{T}\left(\frac{\Gl_j t T}{2\pi}-k_\nmin\right)\right)}  \nonumber \\
&\geqslant \frac{\sin^2\left(\frac{\Gl_j t T}{2}\right)}{\pi^2 k_\nmin^2} \cdot \frac{\frac{\Gl_j t T}{2\pi}+k_\nmin}{\frac{\Gl_j t T}{2\pi}-k_\nmin}
\geqslant
\sin^2\left( \frac{\Gl_jtT}{2}\right) \left(\frac{2\kappa'}{tT} \right)^2
\,.
\end{align}
All the other contributions of the splitting \eqref{sum.split1} are orders $\Ord\big(\ln(\Gl_j t T)/(\Gl_j t T)\big) $. We then get the lower bound in equation \eqref{eq:Gl_square}.


\bigskip

{\bf Acknowledgements.} We thank Seth Lloyd and Giacomo De Palma for useful discussions.


\end{document}